\documentclass[aps,twocolumn,showpacs,superscriptaddress]{revtex4}
\usepackage{graphics}
\usepackage{graphicx}

\begin{document}

\title{Conductivity of a graphene strip: width and gate-voltage dependencies}
\author{F.T. Vasko}
\email{ftvasko@yahoo.com}
\affiliation{Institute of Semiconductor Physics, NAS of Ukraine, Pr. Nauki 41, Kiev,
03028, Ukraine}
\author{I. V. Zozoulenko}
\affiliation{Solid State Electronics, ITN, Link\"{o}ping University 601 74, Norrk\"{o}%
ping, Sweden}
\date{\today}

\begin{abstract}
We study the conductivity of a graphene strip taking into account
electrostatically-induced charge accumulation on its edges. Using a local
dependency of the conductivity on the carrier concentration we find that the
electrostatic size effect in doped graphene strip of the width of 0.5 - 3 $%
\mu $m can result in a significant (about 40\%) enhancement of the effective
conductivity in comparison to the infinitely wide samples. This effect
should be taken into account both in the device simulation as well as for
verification of scattering mechanisms in graphene.
\end{abstract}

\maketitle

Description of the electron transport in wide graphene sheets is usually
based on the Boltzmann approach under the assumption of a homogeneous
distribution of the electron density, see Refs. 1-3 for a review. In the
opposite limit of ultranarrow strips (nanoribbons) the transport is
described within the Landauer-Buttiker approach. With increase of the width
of the strip, $w$, the quasiclassical size-dependent effect, when both
in-plane and edge scattering mechanisms are essential, takes place.
In a typical sample, this regime can be realized at $w\sim l_{mfp}\lesssim
0.1$ $\mu $m; here $l_{mfp}$ is the mean free path. Further increase of the
width, when $w$ becomes comparable to the thickness of a dielectric layer, $%
d $, results in an \textit{electrostatically-induced} size-dependent
modifications of longitudinal transport caused by a redistribution of the
in-plane density near edges of the strip. Such the redistribution takes
place in any capacitor under the applied gate voltage $V_{g}$ and is
illustrated in Fig. 1 for different aspect ratios $d/w$. \cite{5} Recently,
it has been demonstrated that a pronounced charge accumulation takes place
at the edges of a graphene strip \cite{6}. It is therefore important to
investigate how this charge accumulation affects the conductivity of
realistic graphene samples.

\begin{figure}[tbp]
\begin{center}
\includegraphics[scale=1.1]{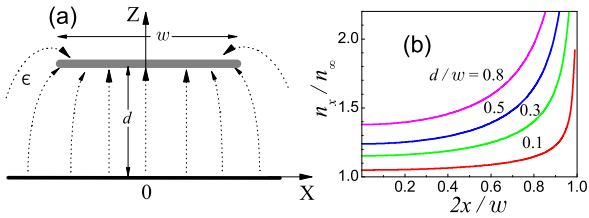}
\end{center}
\addvspace{-1cm}
\caption{(Color online) (a) Cross-section (X0Z-plane) graphene strip of width
$w$ placed at the distance $d$ over a gate. Dashed arrows show the field distribution
under the applied gate voltage $V_{g}$. (b) Planar distribution of the
carrier concentration, $n_{x}$, normalized to the concentration $n_{\infty }$
($n_{\infty }$ is the concentration induced by $V_{g}$ in an infinite
homogeneous structure, $w\to\infty$). For the parameters used ($d=300$ nm
and $\protect\epsilon=3.9$ appropriate for SiO$_{2}$ substrate) here $%
n_{\infty }=\protect\alpha V_{g}$, with $\protect\alpha =7.2\times 10^{10}$cm%
$^{-2}$/V. Ratios $d/w$ are marked.}
\end{figure}

In this paper, we calculate the low-temperature conductivity of a graphene
strip within the local approximation when the mean free path $l_{mfp}$ is
less than the scale of the concentration inhomogeneity on the boundary
(which is $\sim d)$. For the sake of simplicity we restrict ourselves to the
case of the uniform dielectric permittivity, $\epsilon$, i.e. we consider
a structure with cup layer (the effect should increase for the
SiO$_{2}$-graphene structure, see \cite{6}).
A current of the density $J$ flows in the 0Y-direction, see
Fig. 1a. Because of the redistribution of the in-plane
density along 0X-direction near the edges as discussed above, the local
conductivity and the electric field are dependent on the transverse
coordinate such that $J=\sigma (x)E(x)$. Since $J=$ const due to the
continuity equation $\partial J/\partial x=0$, we obtain the
average electric field
\begin{equation}
\left\langle E\right\rangle =\frac{1}{w}\int_{-w/2}^{w/2}E(x)dx=\frac{J}{w}%
\int_{-w/2}^{w/2}\frac{dx}{\sigma (x)}.
\end{equation}
Introducing the relation between $J$ and $\left\langle E\right\rangle $
through the effective conductivity, $\sigma _{eff}$, we obtain
\begin{equation}
\sigma _{eff}=\frac{J}{\left\langle E\right\rangle }=w\left(
\int_{-w/2}^{w/2}\frac{dx}{\sigma (x)}\right) ^{-1}.
\end{equation}%
Below we calculate $\sigma _{eff}$ using the concentration distribution for
a capacitor \cite{5} (see Fig. 1b and Ref. 4b for details of calculations)
and the local quasiclassical conductivity $\sigma (x)$ which depends on the
density $n_{x}$ parametrically.

\begin{figure}[tbp]
\begin{center}
\includegraphics[scale=0.9]{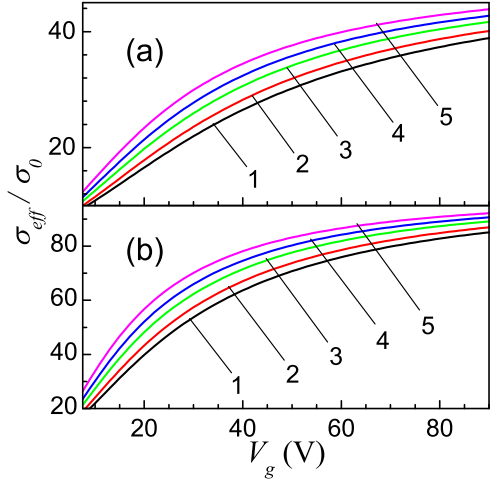}
\end{center}
\addvspace{-1cm}
\caption{(Color online) Effective conductivity of the graphene strip versus
gate voltage for the case I (scattering by finite- and short-range disorder
potentials characterized by $\protect\upsilon_{d}$ and $\protect\upsilon_{0}$%
, respectively) for $\protect\upsilon _{d}/\protect\upsilon $=0.3 and: (a) $%
l_{c}=$6 nm and $\protect\upsilon _{0}/ \protect\upsilon $ =0.02, (b) $%
l_{c}= $9 nm and $\protect\upsilon _{0}/\protect\upsilon $=0.01. Ratios $%
d/w= $0 (1), 0.1 (2), 0.3 (3), 0.5 (4), and 0.8 (5) are marked. }
\end{figure}

For the zero-temperature case, the local conductivity $\sigma (x)$ can be
written through the $x$-dependent Fermi momentum, $p_{x}=\hbar \sqrt{\pi
n_{x}}$, as follows (see \cite{1,2,3,7,8} for details):
\begin{equation}
\sigma (x)=\sigma _{0}\frac{\upsilon p_{x}}{\nu (p_{x})},~~~~\sigma _{0}=%
\frac{e^{2}}{\pi \hbar }.
\end{equation}
Here $\upsilon \simeq 10^{8}$ cm/s, $\sigma _{0}^{-1}\simeq $12.9 k$\Omega $,
and $\nu (p)$ is the momentum relaxation rate. In order to
check a sensitivity of the size effect under consideration to scattering
details, we consider here the momentum relaxation rates caused by: (I)
Gaussian and short-range disorder potentials, \cite{7,8} with the total
relaxation rate $\nu _{G}(p)+\nu _{S}(p)$, and (II) Coulomb and short-range
disorder potentials, \cite{3} with the total rate $\nu _{C}(p)+\nu _{S}(p)$.
The Gaussian disorder is described by the potential $W_{\mathbf{x}}$ with
the Gaussian correlation function $\left\langle W_{\mathbf{x}}W_{\mathbf{x}%
^{\prime }}\right\rangle =\overline{W}^{2}\exp \left[ -(\mathbf{x}-\mathbf{x}%
^{\prime })^{2}/2l_{c}^{2}\right] $, where $\overline{W}$ is the averaged
energy and $l_{c}$ is the correlation length. Within the Born approximation,
the correspondent relaxation rate reads $\nu _{G}(p)=(\upsilon _{d}p/\hbar
)\Psi (pl_{c}/\hbar )$ where we have introduced the dimensionless function
$\Psi (z)=e^{-z^{2}}I_{1}(z^{2})/z^{2}$ with the first-order Bessel function
of an imaginary argument, $I_{1}(z)$ and the characteristic velocity
$\upsilon_{d}=\pi (\overline{W}l_{c}/\hbar )^{2}/(2\upsilon )$.
The relaxation rate due to the short-range disorder potential within the
Born approximation
has a similar (if $l_{c}\rightarrow 0$) dependence $\nu _{S}(p)=\upsilon
_{0}p/\hbar $, with an explicit expression for the characteristic velocity
$\upsilon _{0}$
given in Refs. 6b and 7a. Note that in the present study we limit ourselves to the case
of weak short-range scattering, as opposed to the case of strong
resonant short-range scattering where $\nu _{S}(p)$ has a different functional dependence
and can not be described within the Born approximation [6b, 7b]. Since our aim is to
demonstrate that the size-effect under consideration is generic for graphene strips and
takes place for different scattering conditions, a comparative discussion of scattering
mechanisms is beyond the scope of this work. The momentum relaxation rate due to
scattering by charged Coulomb impurities of concentration $n_{im}$ is given by \cite{3}
$\nu _{C}(p)=n_{im}(\pi e^{2}/\epsilon )^{2}/(\hbar
\upsilon p)$,
where the screening effect is omitted (screening reduces $\nu _{C}$ but does
not change the relation $\nu _{C}\propto p^{-1}$).

\begin{figure}[tbp]
\begin{center}
\includegraphics[scale=0.9]{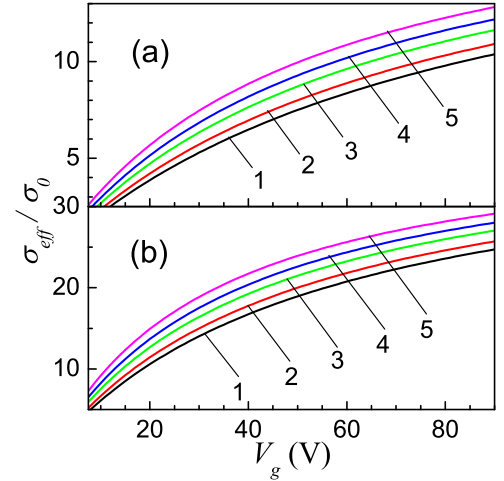}
\end{center}
\addvspace{-1cm}
\caption{(Color online) The same as in Fig. 2 for the case II (scattering by
charge impurities and short-range disorder) for: (a) $n_{im}=3\times 10^{11}$
cm$^{-2}$ and $\protect\upsilon _{0}/\protect\upsilon $=0.05, (b) $%
n_{im}=10^{11}$ cm$^{-2}$ and $\protect\upsilon _{0}/\protect\upsilon $%
=0.025.}
\end{figure}

Using these relaxation rates in Eq. (3) we obtain for the local conductivity
for the cases (I) and (II),
\begin{equation}
\sigma (x)=\sigma _{0}\upsilon \left\{ \begin{array}{*{20}c} \left[
{\upsilon \Psi \left( {\sqrt {\pi l_c n_x } } \right) + \upsilon _0 }
\right]^{ - 1} , & (I) \\ \left( {\upsilon \Gamma n_{im} /n_x + \upsilon _0
} \right)^{ - 1} , & (II) \\ \end{array}\right.
\end{equation}%
where $\Gamma =\pi (e^{2}/\varepsilon \hbar \upsilon )^{2}$. Further,
substituting (4) into Eq. (2) and performing numerical integration we obtain
the effective conductivity $\sigma _{eff}$ versus $V_{g}$ and $w$ for the
cases (I) and (II).

\begin{figure}[tbp]
\begin{center}
\includegraphics{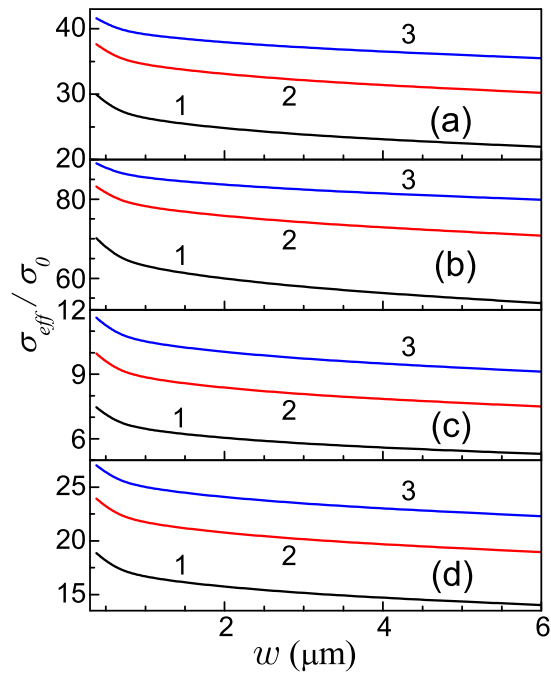}
\end{center}
\addvspace{-1cm}
\caption{(Color online) Effective conductivity $\protect\sigma _{eff}/%
\protect\sigma _{0}$ versus width $w$ for $V_{g}=$30 V (1), 50 V (2), and 70
V (3) at conditions used in: (a) Fig. 2a, (b) Fig. 2b, (c) Fig. 3a, and (d)
Fig. 3b.}
\end{figure}

Figure 2 shows the gate voltage dependence for $\sigma _{eff}/\sigma _{0}$
for the case (I) for different $l_{c}$, $\upsilon _{0}/\upsilon $, and
relative widths, $d/w$. The ratio $\upsilon_{d}/\upsilon $=0.3 is used and
these parameters are correspondent to a sample with maximal resistance per
square $\sim $2 k$\Omega$). In Fig. 3 the same dependencies are presented
for the case (II) for the values of $n_{im}$ and $\upsilon _{0}/\upsilon $
corresponding to the maximal resistance $\sim $4 k$\Omega$. Since the local
conductivity (4) increases with the carrier concentration (at $V_{g}\sim $%
100 V the saturation due to the short-range scattering takes place), $\sigma
_{eff}/\sigma _{0}$ increases with $d/w$ in agreement with the concentration
dependency shown in Fig. 1b. A decrease of $\sigma _{eff}$ with increase of
the width of the strip, $w$, is shown in Fig. 4 for the cases presented in
Figs. 2 and 3 for different $V_{g}$. For $w\approx d$ the effective
conductivity can exceed the corresponding conductivity for the infinite
sample up to 50\%. For a more narrow strip, $\sigma _{eff}$ shows even
stronger increase; however it is not apparent that the local approximation
is valid in realistic systems when $w<d$.

Next, we briefly discuss the reported transport measurements in graphene
\cite{1,2,3,9,10,11,12,13} in relation to our results concerning the
electrostatically-induced size-dependent modifications of the longitudinal
conductivity. Only in a few cases, \cite{11} the condition $w\gg d$ was
satisfied and the size effect under consideration was not essential. In some
papers, \cite{12} a situation is not clear due to the effect of side gate
or/and additional contacts along a conducting canal. In other measurements,
e.g. in \cite{13}, the dependency $\sigma_{eff}(V_{g})$ is expected to be
affected due to the electrostatic size effect. However, a verification of a
size effect through $V_{g}$ dependencies is uncertain because these
dependencies are similar for the cases (I) and (II) above. In addition, a
microscopic mechanism of scattering in graphene remains under debate. \cite%
{3,14} Direct measurements of the size-dependent conductivity can be
performed using a set of samples of different widths made from the same
flake (to avoid contact effect 
these samples should be long enough) or with a multi-contacted sample formed
by fragments of different widths, similar to those used in Ref. 10b. Note
also a possibility for direct STM measurements of a lateral charge
distribution. In addition, the quasiclassical magnetotransport should be
modified if $d/w\gtrsim $0.1.

Let us finally list the assumptions used in our calculations. The local
approach is valid when the characteristic scale of a charge redistribution ($%
\sim d$) exceeds the mean free path ($l_{mfp}\sim \upsilon /\overline{\nu }$%
, $\overline{\nu }^{-1}$ is the relaxation time and $l_{mfp}\leq 0.1\mu $m
for a typical case; for the non-local regime, at $l_{mfp}\sim d$, a size
dependency should be modified, but not suppressed). Here we consider a
simple capacitor geometry with homogeneous $\epsilon $; more complicate
calculations are necessary to account for discontinuity of $\epsilon $ on
the dielectric/air interface. Finally, we limited ourselves to the standard
regime of quasiclassical transport (the Boltzmann equation approach and no
electron/hole puddles formation at low $V_{g}$) and we do not consider a
temperature dependency of conductivity.

In summary, we have demonstrated that the electrostatic size effect in doped
graphene strips of width 0.5 - 3 $\mu $m results in a visible (about 40\%)
enhancement of the effective conductivity. This should be taken into account
both for device simulation and for verification of scattering mechanisms in
graphene.

F.T.V. acknowledges support from the Swedish Institute. I.V.Z. acknowledges
support from the Swedish Research Council (VR).


\begin{thebibliography}{99}
\bibitem{1} A. H. Castro Neto, F. Guinea, N.M.R. Peres, K.S. Novoselov, and
A.K. Geim, Rev. Mod. Phys. \textbf{81}, 109 (2009).

\bibitem{2} D. S. L. Abergel, V. Apalkov, J. Berashevich, K. Ziegler, T.
Chakraborty, Adv. Phys. \textbf{59}, 261 (2010); E. R. Mucciolo and C. H.
Lewenkopf, J. Phys.: Condens. Matter \textbf{22}, 273201 (2010).

\bibitem{3} S. Adam, E.H. Hwang, E. Rossi, S. Das Sarma, Solid State Comm.
\textbf{149}, 1072 (2009); S. Das Sarma, S. Adam, W. H.Hwang, and E. Rossi,
arXiv:1003.4731.

\bibitem{5} P. M. Morse and H. Feshbach, \textit{Methods of Theoretical
Physics}, Part II (McGraw-Hill, New York, 1963);
H. Nishiyama and M. Nakamura, IEEE Trans. on Components, Hybrides, and
Manufacting Technology, \textbf{13}, 417 (1990).

\bibitem{6} P. G. Silvestrov and K. B. Efetov, Phys. Rev. B \textbf{77},
155436 (2008); A. A. Shylau, J. W. Klos and I. V. Zozoulenko, Phys. Rev. B
\textbf{80}, 205402 (2009)

\bibitem{7} N. M. R. Peres, J. M. B. Lopes dos Santos, and T. Stauber, Phys.
Rev. B \textbf{76}, 073412 (2007); T. Stauber, N. M. R. Peres, and F.
Guinea. Phys. Rev. B \textbf{76}, 205423 (2007).

\bibitem{8} F.T. Vasko and V. Ryzhii, Phys. Rev. B \textbf{76}, 233404
(2007); J. W. Klos and I. V. Zozoulenko, Phys. Rev. B, in press (2010).

\bibitem{9} K. S. Novoselov, A. K. Geim, S. V. Morozov, D. Jiang, Y. Zhang,
S. V. Dubonos, I. V. Grigorieva, and A. A. Firsov, Science \textbf{306}, 666
(2004); A. K. Geim and K. S. Novoselov, Nat. Mater. \textbf{6}, 183 (2007).

\bibitem{10} Y. Zhang, Y.-W. Tan, H. L. Stormer, and P. Kim, Nature \textbf{%
438}, 201 (2005); Y.-W. Tan, Y. Zhang, H. L. Stormer, and P. Kim, Eur. Phys.
J. Spec. Top. \textbf{148}, 15 (2007).

\bibitem{11} Y.-W. Tan, Y. Zhang, K. Bolotin, Y. Zhao, S. Adam, E. H. Hwang,
S. Das Sarma, H. L. Stormer, and P. Kim, Phys. Rev. Lett. \textbf{99},
246803 (2007); C. Jang, S. Adam, J.-H. Chen, E. D. Williams, S. Das Sarma,
and M. S. Fuhrer, Phys. Rev. Lett. \textbf{101}, 146805 (2008).

\bibitem{12} X. Hong, K. Zou, and J. Zhu, Phys. Rev. B \textbf{80}, 241415
(2009); W. Zhu, V. Perebeinos, M. Freitag, Ph. Avouris, Phys. Rev. B \textbf{%
80}, 235402 (2009).

\bibitem{13} K. S. Novoselov, A. K. Geim, S. V. Morozov, D. Jiang, M. I.
Katsnelson, I. V. Grigorieva, S. V. Dubonos, and A. A. Firsov, Nature
\textbf{438}, 197 (2005). 

\bibitem{14} L. A. Ponomarenko, R. Yang, T. M. Mohiuddin, M. I. Katsnelson,
K. S. Novoselov, S. V. Morozov, A. A. Zhukov, F. Schedin, E. W. Hill, and A.
K. Geim, Phys. Rev. Lett. \textbf{102}, 206603 (2009).

\end{thebibliography}
\end{document}